\documentclass[a4paper,11pt]{article}

\usepackage{enumitem}
\usepackage{float}
\usepackage[margin=2.5cm]{geometry}
\usepackage[grumpy]{gitinfo}
\usepackage{graphicx}
\usepackage{times}
\usepackage[hyphens]{url}
\usepackage{xspace}

\makeatletter
\g@addto@macro{\UrlBreaks}{\UrlOrds}
\makeatother

\usepackage{amsmath}
\allowdisplaybreaks[2]

\usepackage{amssymb}

\usepackage{amsthm}
\newtheoremstyle{plain-boldhead}
  {\topsep}
  {\topsep}
  {\itshape}
  {}
  {\bfseries}
  {.}
  { }
  {\thmname{#1}\thmnumber{ #2}\thmnote{ (\bfseries #3)}}
\newtheoremstyle{definition-boldhead}
  {\topsep}
  {\topsep}
  {\normalfont}
  {}
  {\bfseries}
  {.}
  { }
  {\thmname{#1}\thmnumber{ #2}\thmnote{ (\bfseries #3)}}
\theoremstyle{plain-boldhead}

\theoremstyle{definition-boldhead}

\newtheorem{remark}{Remark}

\floatstyle{ruled}
\newfloat{algo}{htbp}{algo}
\floatname{algo}{Algorithm}

\newcommand{\str}[1]{\textsc{#1}}
\newcommand{\var}[1]{\textit{#1}}
\newcommand{\op}[1]{\textsl{#1}}

\newcommand{\true}{\str{True}\xspace}

\newcommand{\shortlist}{\parskip0pt\topsep0pt\partopsep0pt\parsep1pt\itemsep2pt}

\newcommand{\Y}{\ensuremath{\,\checkmark}\,}
\newcommand{\YY}{\ensuremath{\checkmark}}
\newcommand{\N}{\ensuremath{-}\xspace}
\newcommand{\E}{\ensuremath{.}}

\begin{document}

\title{\bf Blockchain Consensus Protocols in the Wild}

\author{Christian Cachin \and Marko Vukoli\'c}

\date{IBM Research - Zurich\\
  \url{(cca|mvu)@zurich.ibm.com}\\[2ex]
  \gitCommitterDate}

\maketitle

\begin{abstract}\noindent
  A blockchain is a distributed ledger for recording transactions,
  maintained by many nodes without central authority through a distributed
  cryptographic protocol.  All nodes validate the information to be
  appended to the blockchain, and a consensus protocol ensures that the
  nodes agree on a unique order in which entries are appended.  Consensus
  protocols for tolerating Byzantine faults have received renewed attention
  because they also address blockchain systems.  This work discusses the
  process of assessing and gaining confidence in the resilience of a
  consensus protocols exposed to faults and adversarial nodes.  We advocate
  to follow the established practice in cryptography and computer security,
  relying on public reviews, detailed models, and formal proofs; the
  designers of several practical systems appear to be unaware of this.
  Moreover, we review the consensus protocols in some prominent
  permissioned blockchain platforms with respect to their fault models and
  resilience against attacks.  The protocol comparison covers Hyperledger
  Fabric, Tendermint, Symbiont, R3~Corda, Iroha, Kadena, Chain, Quorum,
  MultiChain, Sawtooth Lake, Ripple, Stellar, and IOTA.
\end{abstract}

\pagestyle{plain}

\section{Introduction}

\emph{Blockchains} or \emph{distributed ledgers} are systems that provide a
trustworthy service to a group of \emph{nodes} or parties that do not fully
trust each other.  They stand in the tradition of distributed protocols for
secure multiparty computation in cryptography and replicated services
tolerating Byzantine faults in distributed systems.  Blockchains also
contain many elements from cryptocurrencies, although a blockchain system
can be conceived without a currency or value tokens.  Generally, the
blockchain acts as a trusted and dependable third party, for maintaining
shared state, mediating exchanges, and providing a secure computing engine.
Many blockchains can execute arbitrary tasks, typically called
\emph{smart contracts}, written in a domain-specific or a general-purpose
programming language.

In a \emph{permissionless blockchain}, such as Bitcoin or Ethereum, anyone can be a
user or run a node, anyone can ``write'' to the shared state through
invoking transactions (provided transaction fees are paid for), and anyone
can participate in the consensus process for determining the ``valid''
state.  A \emph{permissioned blockchain} in contrast, is operated by known
entities, such as in \emph{consortium blockchains}, where members of a
consortium or stakeholders in a given business context operate a
permissioned blockchain network. Permissioned blockchains systems have
means to identify the nodes that can control and update the shared state,
and often also have ways to control who can issue transactions.  A
\emph{private blockchain} is a special permissioned blockchain operated by
one entity, i.e., within one single trust domain.

Permissioned blockchains address many of the problems that have been
studied in the field of distributed computing over decades, most
prominently for developing \emph{Byzantine fault-tolerant (BFT)} systems.
Such blockchains can benefit from many techniques developed for reaching
consensus, replicating state, broadcasting transactions and more, in
environments where network connectivity is uncertain, nodes may crash or
become subverted by an adversary, and interactions among nodes are
inherently asynchronous.
The wide-spread interest in blockchain technologies has triggered
new research on
practical distributed consensus protocols.  There is also a growing number
of startups, programmers, and industry groups developing blockchain
protocols based on their own ideas, not relying on established knowledge.

The purpose of this paper is to give an overview of consensus protocols
actually being used in the context of permissioned blockchains, to review
the underlying principles, and to compare the resilience and
trustworthiness of some protocols.  We leave out permissionless (or
``public'') blockchains that are coupled to a cryptocurrency and their
consensus protocols, such as \emph{proof-of-work} or \emph{proof-of-stake},
although this is a very interesting subject by itself.

We start by pointing out that developing consensus protocols is difficult
and should not be undertaken in an ad-hoc manner (Sec.~\ref{sec:trust}).  A
resilient consensus protocol is only useful when it continues to deliver
the intended service under a wide range of adversarial influence on the
nodes and the network.  Detailed analysis and formal argumentation are
necessary to gain confidence that a protocol achieves its goal.  In that
sense, distributed computing protocols resemble cryptosystems and other
security mechanisms; they require broad agreement on the underlying
assumptions, detailed security models, formal reasoning, and wide-spread
public discussion.  Any claim for a ``superior'' consensus protocol that
does not come with the necessary formal justification should be dismissed,
analogously to the approach of ``security by obscurity,'' which is
universally rejected by experts.

The text continues in Section~\ref{sec:consensus} with a brief review of
consensus in the context of blockchains.  As an example for the principle
of scientific study, some serious flaws in a BFT consensus protocol called
\emph{Tangaroa} (also known as \emph{BFT-Raft}) are shown. This protocol
has gained quite some popularity among proponents of permissioned
blockchains but it is not implemented in any blockchain platform.

In Section~\ref{sec:perm} the consensus protocols in a number of
permissioned blockchain platforms are compared, based on the available
product descriptions or source code.  Section~\ref{sec:public} discusses
the consensus mechanisms of blockchain platforms not directly following the
BFT approach: \emph{Sawtooth Lake}, \emph{Ripple} and \emph{Stellar}, and
the \emph{IOTA Tangle}.  A summary concludes the paper.

\section{Trust in a blockchain protocol}
\label{sec:trust}

A blockchain or a distributed ledger protocol can be summarized as a secure
distributed protocol that achieves certain task satisfying targeted
consistency (typically atomicity or total order) and
liveness (or availability) semantics. However, verifying and establishing
trust in the security, consistency and liveness semantics of a blockchain
protocol is challenging.  This section argues that developing consensus
protocols is similar to building cryptographic systems and that it should
benefit from the experience and practice in cryptography.

It is generally difficult to evaluate any security mechanism.  First and
foremost, a ``secure'' solution should not interfere ``too much'' with the
functionality, i.e., primary task one tries to accomplish.  But more
importantly, the security tool should ensure that one can accomplish this
task in a way that is resilient to problems caused by the (adversarial)
environment, by preventing, deterring, withstanding, or tolerating any
influence that could hinder one from accomplishing the task.

Showing that the solution works in the absence of problems and attacks is
easy.  The task is achieved, and output is straightforward to verify.
Assessing the security is the hard part.  A security solution should come
with a clearly stated security model and trust assumption, under which the
solution should satisfy its goal.  This is widely accepted today; it
prompts the question of how to validate that the solution satisfies its
goal.

Yet, the experimental validation of a security solution in the information
technology space fails very often because no experiment can exhaustively
test the solution in all scenarios permitted by the model.  In a way,
experimentation can only demonstrate the \emph{failure} of a security
mechanism.

Therefore one needs to apply mathematical reasoning and formal tools to
reason why the solution would remain secure under \emph{any} scenario
permitted by the stated trust assumption.  Without such reasoning,
security claims remain vague.

In the domain of blockchain protocols, one can learn a lot from the history
of cryptography.  Already since the 19th century, Kerckhoffs' principle has
been widely accepted, which states that ``a cryptosystem should be secure
even if everything about the system, except the key, is public knowledge.''
It implies that any security claim of the kind that a system embodies a
superior but otherwise undisclosed design should be dismissed immediately.

Starting from the pioneering work in the 1980s, modern cryptography has
developed formal treatment, security notions, and corresponding provably
secure protocols.
Cryptography research has concentrated on mathematically formalizing (a
small number of) security assumptions, such as ``computing discrete
logarithms in particular groups is hard,'' and on building complex systems
and protocols that rely on these assumptions, without introducing any
additional insecurity.  In other words, in a ``provably secure'' solution,
an attack on the stated goal of the solution can be turned efficiently into
a violation of some underlying assumption.

For assessing whether the formal models are appropriate and whether the
security assumptions cover the situation encountered during deployment,
human judgment is needed, best exerted through careful review, study,
validation, and expert agreement.  The AES block cipher, for instance, was
selected in 2000 by the U.S. NIST after a multi-year public review process
during which many candidates were debated and assessed openly by the
world-wide cryptographic research community.

During the internet boom in the late 1990s there were many claims of new
and ``unbreakable'' cryptosystems, all lacking substantiation.  Many of
them were covered by Schneier in blog posts about ``snake oil,'' alluding
to the history of medicine before regulation~\cite{schnei99}:
\begin{quote}
  The problem with bad security is that it looks just like good
  security. You can't tell the difference by looking at the finished
  product. Both make the same security claims; both have the same
  functionality. (\dots) Both might use the same protocols, implement the same
  standards, and have been endorsed by the same industry groups. Yet one is
  secure and the other is insecure.
\end{quote}
Expert judgment, formal reasoning, experience, public discussion, and open
validation are needed for accepting a cryptosystem as secure.

A similar development has taken place with building resilient distributed
systems, whose goal is to deliver a service while facing network outages,
communication failures, timing uncertainty, power loss and more.  The
Chubby database of Google~\cite{chgrre07} and Yahoo!'s
ZooKeeper~\cite{hkjr10}, developed for synchronizing critical configuration
information across data centers, support strong consistency and high
availability through redundancy and tolerate benign failures and network
outages.  Those systems started from well-understood, mathematically
specified, and formally verified protocols in the research literature
(e.g., Lamport's Paxos protocol~\cite{lampor98}).  Yet it has taken
considerable effort during development and testing and frequent exercising
of failure scenarios during deployment~\cite{JonesPM16} to achieve the
desired level of resilience in practice.

Over the recent years countless proposals for new features in distributed
ledger systems and completely new blockchain protocols have appeared,
mainly originating from the nascent fintech industry and startup scene.
Most of them come without formal expression of their trust assumption and
security model.  There is no agreed consensus in the industry on which
assumptions are realistic for the intended applications, not to mention any
kind of accepted standard or validation for protocols.
The field of blockchain protocols is in its infancy today, but already
appears at the peak of overstated expectations~\cite{gartne16}.  Many
fantastic and bold claims are made in the fintech and blockchain space by
startups, established companies, researchers, and self-proclaimed experts
alike.  This creates excitement but also confusion in the public opinion.  

Broad agreement on trust assumptions, security models, formal reasoning
methods, and protocol goals is needed.  Developers, investors, and users in
the industry should look towards the established scientific methodology in
cryptography and security with building trustworthy systems, before they
entrust financial value to new protocols.  Open discussion, expert
reviews, broad validation, and standards recommendations should take over
and replace the hype.

\section{Consensus}
\label{sec:consensus}

This section presents background and models for consensus in permissioned
blockchains, first introducing the underlying concept of state-machine
replication in Section~\ref{ssec:smr}.  Sections~\ref{ssec:crashcons} and
\ref{ssec:byzcons} briefly review the most prominent family of protocols
for this task, which is based Paxos/Viewstamped Replication (VSR) and PBFT.
The essential step of transaction validation is discussed in
Section~\ref{ssec:validity}.  To round off this part, we demonstrate the
pitfalls of consensus-protocol design in Section~\ref{ssec:tangaroa}, by
analyzing a proposed BFT consensus protocol called \emph{Tangaroa} and
showing that it does not achieve its goals.

\subsection{Blockchains and consensus}
\label{ssec:smr}

A \emph{blockchain} is a distributed database holding a continuously
growing list of records, controlled by multiple entities that may not trust
each other.  Records are appended to the blockchain in batches or
\emph{blocks} through a distributed protocol executed by the nodes powering
the blockchain.  Each block contains a cryptographic hash of the previous
block, which fixes all existing blocks and embeds a secure representation
of the complete chain history into every block. Additional integrity
measures are often used in potentially malicious, Byzantine environments,
such as the requirement that a block hash is smaller than a given target
(e.g., in Nakamoto-style proof-of-work consensus), or a multi-signature (or
a threshold signature) over a block, by the nodes powering the blockchain
(for permissioned blockchains).  The nodes communicate over a network and
collaboratively construct of the blockchain without relying on a central
authority.

However, individual nodes might crash, behave maliciously, act against the
common goal, or the network communication may become interrupted.  For
delivering a continuous service, the nodes therefore run a fault-tolerant
\emph{consensus protocol} to ensure that they all agree on the order in
which entries are appended to the blockchain.

Since the whole blockchain acts as a trusted system, it should be
\emph{dependable}, \emph{resilient}, and \emph{secure}, ensuring properties
such as availability, reliability, safety, confidentiality, integrity and
more~\cite{alrl04}.  A blockchain protocol ensures this by
\emph{replicating} the data and the operations over many nodes.
Replication can have many
roles~\cite{schnei90,CharronBostPS10,Kleppmann17}, but blockchains
replicate data only for resilience, not for scalability.  All nodes
validate, in principle, the information to be appended to the blockchain;
this feature stimulates the trust of all nodes in that the blockchain as
a whole operates correctly.

For assessing a blockchain protocol, it is important to be clear about the
underlying \emph{trust assumption} or \emph{security model}.  This
specifies the environment for which the protocol is designed and in which
it satisfies its guarantees.  Such assumptions should cover all elements in
the system, including the network, the availability of synchronized clocks,
and the expected (mis-)behavior of the nodes.  For instance, the typical
generic trust assumption for a system with $n$ independent nodes says that
no more than $f < n/k$ nodes become \emph{faulty} (crash, leak information,
perform arbitrary actions, and so on), for some $k = 2, 3, \dots$.  The
other $n-f$ nodes are \emph{correct}.  A trust assumption always represents
an idealization of the real world; if some aspect not considered by the
model can affect the actually deployed system, then the security must be
reconsidered.

\paragraph{State-machine replication.}
The formal study and development of algorithms for exploiting replication
to build resilient servers and distributed systems goes back to Lamport et
al.'s pioneering work introducing Byzantine
agreement~\cite{peshla80,lashpe82}.  The topic has evolved through a long
history since and is covered in many
textbooks~\cite{AttiyaW04,CachinGR11,Raynal10b,Vukolic12}; a good summary
can be found in a ``30-year perspective on
replication''~\cite{CharronBostPS10}.

As summarized concisely by Schneider~\cite{schnei90}, the task of reaching
and maintaining consensus among distributed nodes can be described with two
elements: (1) a (deterministic) \emph{state machine} that implements the
logic of the service to be replicated; and (2) a \emph{consensus protocol}
to disseminate requests among the nodes, such that each node executes the
\emph{same sequence of requests} on its instance of the service.  In the
literature, ``consensus'' means traditionally only the task of reaching
agreement on one single request (i.e., the first one), whereas ``atomic
broadcast''~\cite{hadtou93} provides agreement on a sequence of requests,
as needed for state-machine replication.  But since there is a close
connection between the two (a sequence of consensus instances provides
atomic broadcast), the term ``consensus'' more often actually stands for
atomic broadcast, especially in the context of blockchains.  We adopt this
terminology here and also use ``transaction'' and ``request'' as synonyms
for one of the messages to be delivered in atomic broadcast.

\paragraph{Asynchronous and eventually synchronous models.}
Throughout this text, we assume the \emph{eventual-synchrony} network
model, introduced by Dwork et al.~\cite{dwlyst88}.  It models an
asynchronous network that may delay messages among correct nodes
arbitrarily, but eventually behaves synchronously and delivers all messages
within a fixed (but unknown) time bound.  Protocols in this model never
violate their consistency properties (safety) during asynchronous periods,
as long as the assumptions on the kind and number of faulty nodes are met.
When the network stabilizes and behaves synchronously, then the nodes are
guaranteed to terminate the protocol (liveness).  Note that a protocol may
stall during asynchronous periods; this cannot be avoided due to a
fundamental discovery by Fischer et al.~\cite{filypa85} (the celebrated
``FLP impossibility result''), which rules out that deterministic protocols
reach consensus in (fully) asynchronous networks.

The model is widely accepted today as realistic for designing resilient
distributed systems.  Replication protocols have to cope with network
interruptions, node failures, system crashes, planned downtime, malicious
attacks by participating nodes, and many more unpredictable effects.
Developing protocols for asynchronous networks therefore provides the best
possible resilience and avoids any assumptions about synchronized clocks
and timely network behavior; making such assumptions can quickly turn into
a vulnerability of the system if any one is not satisfied during
deployment.

Protocol designers today prefer the \emph{eventual synchrony} assumption
for its simplicity and practitioners observe that it has broader coverage
of actual network behavior, especially when compared to so-called partially
synchronous models that assume probabilistic network behavior over time.

\paragraph{Consensus in blockchain.}
Although Nakamoto's Bitcoin paper~\cite{nakamo09} does not explicitly
mention the \emph{state-machine replication} paradigm~\cite{schnei90},
Bitcoin establishes consensus on one shared ledger based on voting among
the nodes: ``(Nodes) vote with their CPU power, expressing their acceptance
of valid blocks by working on extending them and rejecting invalid blocks
by refusing to work on them.  Any needed rules and incentives can be
enforced with this consensus mechanism''~\cite{nakamo09}.

With the work of Garay et al.~\cite{gakile15}, a formal equivalence between
the task solved by the ``Nakamoto protocol'' inside Bitcoin and the
consensus problem in distributed computing was shown for the first time.
This result coincided with the insight, developed in the fintech industry,
that a blockchain platform may use a generic consensus mechanism and
implement it with any protocol matching its trust model~\cite{swanso15}.
In today's understanding a blockchain platform may use an arbitrary
consensus mechanism and retain most of its further aspects like
distribution, cryptographic immutability, and transparency.

Existing consensus and replication mechanisms have therefore received
renewed attention, for applying them to blockchain systems.  Several
protocols relevant for blockchains are reviewed in the next sections.  We
discuss only protocols for static groups here; they require explicit group
reconfiguration~\cite{srmj12,besoal14} and do not change membership
otherwise.  This assumption contrasts with view-synchronous
replication~\cite{chkevi01}, where the group composition may change
implicitly by removing nodes perceived as unavailable.

\subsection{Crash-tolerant consensus}
\label{ssec:crashcons}

As mentioned earlier, the form of consensus relevant for blockchain is
technically known as \emph{atomic broadcast}.  It is formally obtained as
an extension of a \emph{reliable broadcast} among the node, which also
provides a global or \emph{total order} on the messages delivered to all
correct nodes.  An atomic broadcast is characterized by two (asynchronous)
events \op{broadcast} and \op{deliver} that may occur multiple times.
Every node may broadcast some message (or transaction) $m$ by invoking
$\op{broadcast}(m)$, and the broadcast protocol outputs $m$ to the local
application on the node through a $\op{deliver}(m)$ event.

Atomic broadcast ensures that each correct node outputs or \emph{delivers}
the same sequence of messages through the \op{deliver} events.  More
precisely~\cite{hadtou93,CachinGR11}, it ensures these properties:
\begin{description}\shortlist
\item[Validity:] If a correct node~$p$ broadcasts a message~$m$, then $p$
  eventually delivers~$m$.
\item[Agreement:] If a message~$m$ is delivered by some correct node,
  then $m$ is eventually delivered by every correct node.
\item[Integrity:] No correct node delivers the same message more than once;
  moreover, if a correct node delivers a message~$m$ and the sender~$p$ of
  $m$ is correct, then $m$ was previously broadcast by~$p$.
\item[Total order:] For messages $m_1$ and $m_2$, suppose $p$ and $q$ are
  two correct nodes that deliver $m_1$ and $m_2$.  Then $p$ delivers $m_1$
  before $m_2$ if and only if $q$ delivers $m_1$ before $m_2$.
\end{description}

The most important and most prominent way to implement atomic broadcast
(i.e., consensus) in distributed systems prone to $t < n/2$ node crashes is
the family of protocols known today as \emph{Paxos}~\cite{lampor98,lampor01}
and \emph{Viewstamped Replication (VSR)}~\cite{okilis88,liscow12}.
Discovered independently, their core mechanisms exploit the same
ideas~\cite{lampso01,liskov10}.  They have been implemented in dozens of
mission-critical systems and power the core infrastructure of major cloud
providers today~\cite{chgrre07}.

The \emph{Zab} protocol inside \emph{ZooKeeper} is a prominent member of
the protocol family; originally from Yahoo!, it is available as open
source~\cite{hkjr10,jurese11,rescsc15}
(\url{https://zookeeper.apache.org/}) and used by many systems.  A more
recent addition to the family is \emph{Raft}~\cite{ongous14}, a specialized
variant developed with the aim of simplifying the understanding and the
implementation of Paxos.  It is contained in dozens of open-source tools
(e.g., \emph{etcd} -- \url{https://github.com/coreos/etcd}).

All protocols in this family progress in a sequence of \emph{views} or
``epochs,'' with a unique leader for each view that is responsible for
progress.  If the leader fails, or more precisely, if the other nodes
suspect that the leader has failed, they can replace the current leader by
moving to the next view with a fresh leader.  This \emph{view change}
protocol must ensure agreement, such that message already delivered by a
node in the abandoned view is retained and delivered by all correct nodes
in this or another future view.

\subsection{Byzantine consensus}
\label{ssec:byzcons}

More recently, consensus protocols for tolerating \emph{Byzantine} nodes
have been developed, where nodes may be \emph{subverted} by an adversary
and act maliciously against the common goal of reaching agreement.  In the
eventual-synchrony model considered here, the most prominent protocol is
\emph{PBFT (Practical Byzantine Fault-Tolerance)}~\cite{caslis02}.  It can
be understood as an extension of the Paxos/VSR
family~\cite{lampso01,cachin09,liskov10} and also uses a progression of
views and a unique leader within every view.  In a system with $n$ nodes
PBFT tolerates $f < n/3$ Byzantine nodes, which is optimal.  Many research
works have analyzed and improved aspects of it and made it more robust in
prototypes~\cite{cwadm09}.  A proposed extension of the Raft variant of
Paxos/VSR called \emph{Tangaroa} has been analyzed in
Section~\ref{ssec:tangaroa}.

Actual systems that implement PBFT or one of its variants are much harder
to find than systems implementing Paxos/VSR.  In fact, \emph{BFT-SMaRt}
(\url{https://github.com/bft-smart/library}) is the only known project that
was developed before the interest in permissioned blockchains surged around
2015~\cite{swanso15}.  Actually, Bessani et al.~\cite{besoal14,bessou12}
from the University of Lisbon started work on it around~2010.  There is
widespread agreement today that BFT-SMaRt is the most advanced and most
widely tested implementation of a BFT consensus protocol available.
Experiments have demonstrated that it can reach a throughput of about
80'000 transactions per second in a LAN~\cite{besoal14} and very low
latency overhead in a WAN~\cite{soubes15}.

Like Paxos/VSR, Byzantine consensus implemented by PBFT and BFT-SMaRt
expects an eventually synchronous network to make progress.  Without this
assumption, only \emph{randomized} protocols for Byzantine consensus are
possible, such as the practical variations relying on distributed
cryptography~\cite{cakush05} as prototyped by SINTRA~\cite{cacpor02} or,
much more recently, HoneyBadger~\cite{mxcss16}.

\subsection{Validation}
\label{ssec:validity}

In an atomic broadcast protocol resilient to crashes, every message is
usually considered to be an acceptable request to the service.  For
Byzantine consensus, especially in blockchain applications, it makes sense
to ask that only ``valid'' transactions are output by the broadcast
protocol.  To formalize this, the protocol is parameterized with a
deterministic, external predicate~$V()$, such that the protocol delivers
only messages satisfying~$V()$.  This notion has been introduced as
\emph{external validity} by Cachin et al.~\cite{ckps01}.

The predicate must be deterministically computable locally by every process.  More precisely,
$V()$ must guarantee that when two correct nodes~$p$ and $q$ in an atomic
broadcast protocol have both delivered the same sequence of messages up to
some point, then $p$ obtains $V(m) = \true$ for any message~$m$ if and only
if $q$ also determines that~$V(m) = \true$.

This combination of transaction validation and establishing consensus is
inherent in permissionless blockchains based on proof-of-work consensus,
such as Bitcoin and Ethereum.  For permissioned-blockchain protocols, one
could in principle also separate this step from consensus and perform the
(deterministic) validation of transactions on the ordered ``raw'' sequence
output by atomic broadcast.  This could make the protocol susceptible to
denial-of-service attacks from clients broadcasting excessively many
invalid transactions.  Hence most consensus protocols reviewed in this text
combine ordering with validation and use a form of external validity based
on the current blockchain state.

\subsection{Tangaroa (BFT-Raft) is neither live nor safe}
\label{ssec:tangaroa}

Before we address the protocols found in practical blockchain platforms, we
discuss some flaws in a proposed BFT consensus protocol that are typical
for the kind of mistakes one can make.  \emph{Tangaroa}~\cite{copzho14} is
intended as an extension of the \emph{Raft}~\cite{ongous14} consensus
protocol for BFT, developed as a class project in a university course on
distributed systems.  Although it was neither peer-reviewed nor published
in the scientific literature, and for reasons unknown to us, Tangaroa has
become quite prominent as a potential BFT consensus protocol for
permissioned blockchains.
Like the other consensus protocols discussed earlier, \emph{Tangaroa} uses
the eventual-synchrony network model mentioned above, introduced by Dwork
et al.~\cite{dwlyst88}.  It must never violate agreement (safety), no
matter how the network behaves, and a correct leader must ensure liveness
during synchronous periods, i.e., that messages are continuously delivered.

Raft itself represents a simplified member of the Paxos/VSR family of
crash-tolerant replication and consensus protocols; it was developed for
pedagogical reasons~\cite{ongous14}.  Like Paxos/VSR Raft proceeds in
successive \emph{views} and defines the role of a \emph{leader node} for
each view.  This node is responsible for driving the protocol.  When the
protocol within one view stalls, every node by itself will try to become
leader after waiting for a random time.  A leader election phase follows,
from which one node will emerge as leader and be accepted by all other
(follower) nodes.  Once established, the leader orders and replicates
messages within its view.

For extending the structure behind Raft to tolerate $f < n/3$ Byzantine
nodes, one has to solve at least the leader-election problem and the
reliable message replication within a view.  Although it is well-known how
to achieve this, as demonstrated by PBFT~\cite{caslis02}, many related
protocols~\cite{agkqv15}, as well as pedagogical descriptions of
PBFT~\cite{liskov10,CachinGR11}, Tangaroa fails in at least two ways to
withstand maliciously acting nodes:
\begin{description}
\item[\normalfont\em Liveness issue.] If the network is synchronous (i.e.,
  well-connected, messages are delivered timely, and nodes are
  synchronized), then the protocol should be live and continuously order
  messages.  However, since any node may propose itself to become leader, a
  malicious node might simply rush in at the start of a view, become
  elected as leader, and then refuse to perform any further work.  There is
  no way for the nodes to verify that the node waited for its timeout to
  expire.  Hence, the protocol violates liveness.

\item[\normalfont\em Safety issue.] The leader of a view should ensure that
  all correct nodes deliver the same messages in the same order.  This is
  complicated because the leader might fail or even actively try to make
  the nodes disagree on the delivered message.  After a view change, the
  next leader must be able to resume from a consistent state and be sure it
  does not deliver a message twice or omit one.

  However, Tangaroa directly uses the low-level messaging structure from
  Raft, even though it is well-known that additional rounds of exchanges
  are necessary to cope with the problems that a Byzantine leader might
  create.  This can lead to a violation of agreement by Tangaroa, in the
  sense that two correct nodes decide differently; in a blockchain, their
  ledgers would fork and this is a deathblow for a permissioned blockchain.
\end{description}

We illustrate the reason for the safety problem with \emph{Tangaroa} in
Remark~\ref{rem:one}.  Readers not interested in pitfalls of distributed
consensus protocols should skip to the next subsection.

\begin{remark}[Safety violation by Tangaroa]\label{rem:one}
  Within a view, the leader broadcasts an ``AppendEntries'' message
  containing the payload to be replicated to all nodes.  (To ``broadcast''
  means here only that \emph{if} the leader is correct, it sends the same
  point-to-point message to all nodes.)  A node that receives this (for the
  first time at a given index) echoes it by broadcasting
  ``AppendEntriesResponse'' to all.  When a node receives a Byzantine
  quorum of those, it commits the payload (``applies the \dots\ committed
  entries to its state machine'').  This communication primitive is known
  in the literature as \emph{Byzantine consistent broadcast
    (BCB)}~\cite[Sec.~3.10]{CachinGR11}, it goes back to Srikanth and
  Toueg~\cite{sritou87}.  BCB ensures \emph{consistency} in the sense that
  \emph{if} some two correct nodes deliver any payload, then they deliver
  the \emph{same} payload.  BCB does not ensure that any node delivers any
  payload when the leader is faulty.  BCB neither ensures \emph{agreement}
  in the sense that once some correct node has delivered a payload, any
  other correct node will also deliver that.

  It is now possible that in a network with nodes $A, B, C, D$ the current
  leader $D$ is faulty and causes $A$ to deliver and commit some
  payload~$m$~\cite[Fig.~3.11]{CachinGR11}.  $B$ may have echoed $m$ but
  did not deliver it.  Then $D$ causes a leader change to $C$; from this
  moment on, no more messages reach $A$ due to asynchronous network
  behavior and $D$ behaves as if $m$ was never broadcast.  $C$ has no
  knowledge of $m$ and instead picks another payload~$m'$ and delivers
  that.  Then $A$ and $C$ deliver different payloads $m \neq m'$ and
  violate the total-order and agreement properties of consensus
  (Sec.~\ref{ssec:crashcons} and~\ref{ssec:byzcons}).

  It is well-known how to prevent this: PBFT~\cite{caslis02} uses
  \emph{Byzantine reliable broadcast (BRB)} in the place of the BCB in
  Tangaroa (BRB was first formulated by Bracha~\cite{bracha87}).  This
  primitive entails a second all-to-all message exchange, formally ensures
  the agreement property, and, most importantly, makes it possible for a
  subsequent leader to gather enough information so as not to violate the
  consensus protocol's guarantees.
\end{remark}

\subsection{Protocols summary}
\label{ssec:summary}

As an outlook to the following two sections, Table~\ref{tab:summary}
presents a summary of the protocols discussed in the remainder of this
work.

\newcommand*\boxrot{\rotatebox{90}}

\begin{table}[p]
  \centering
  \begin{tabular}{@{} l@{\qquad}c@{\quad}c@{\quad}c@{\quad}c@{\quad}c @{}}
    {\bf\shortstack[l]{Which faults are\\tolerated by a
                 protocol?\vspace*{7ex}}}
    & \boxrot{\bf\shortstack[l]{Special-node\\crash}} 
    & \boxrot{\bf\shortstack[l]{Any $t<n/2$\\nodes crash}}
    & \boxrot{\bf\shortstack[l]{Special-node\\subverted}}
    & \boxrot{\bf\shortstack[l]{Any $f<n/3$\\nodes subverted}}
    \\ \hline 
    Hyperledger Fabric/Kafka & \E & \Y & \E & \N \\ 
    Hyperledger Fabric/PBFT  & \E & \Y & \E & \Y \\ 
    Tendermint               & \E & \Y & \E & \Y \\ 
    Symbiont/BFT-SMaRt       & \E & \Y & \E & \Y \\ 
    R3 Corda/Raft            & \E & \Y & \E & \N \\
    R3 Corda/BFT-SMaRt       & \E & \Y & \E & \Y \\
    Iroha/Sumeragi (BChain)  & \E & \Y & \E & \Y \\
    Kadena/ScalableBFT       &  ? &  ? &  ? &  ? \\
    Chain/Federated Consensus& \N & (\YY) & \N & \N \\
    Quorum/QuorumChain       & \N & (\YY) & \N & \N \\
    Quorum/Raft              & \E & \Y & \E & \N \\
    MultiChain $+$           & \E & \Y & \E & \N \\
    \hline
    Sawtooth Lake/PoET       & $\oplus$  & \Y & $\oplus$ & \N \\
    Ripple                   & $\otimes$ & (\YY) & $\otimes$ & \N \\
    Stellar/SCP              &  ? &  ? &  ? &  ? \\
    IOTA Tangle              &  ? &  ? &  ? &  ? \\ \hline
  \end{tabular}
  \caption{Summary of consensus resilience properties, some of which
    use statically configured nodes with a \emph{special} role.  
    Symbols and notes:
    `\YY' means that the protocol is resilient against the fault and
    `\N' that it is not;
    `\E' states that no such \emph{special node} exists in the
    protocol; `?' denotes that the properties cannot be assessed due to
    lack of information; (\YY) denotes the crash of \emph{other} nodes,
    different from the special node; $+$ MultiChain has non-final
    decisions; $\oplus$ PoET assumes trusted hardware available from
    only one vendor; $\otimes$ Ripple tolerates \emph{one} of the five default
    Ripple-operated validators (special nodes) to be subverted.}
  \label{tab:summary}
\end{table}

\section{Permissioned blockchains}
\label{sec:perm}

This section discusses some notable consensus protocols that are part of
(or have at least been proposed for) the following consortium blockchain
systems:
Hyperledger Fabric;
Tendermint;
Symbiont Assembly;
R3 Corda;
Iroha;
Kadena;
Chain;
Quorum; and
MultiChain.
We assume there are $n$ distributed nodes responsible for consensus, but
some systems contain further nodes with other roles.  Each of the
subsections contains a table summarizing the consensus resilience
properties.

\subsection{Overview}

Among the recent flurry of blockchain-consensus protocols, many have not
progressed past the stage of a paper-based description.  In this section,
we review only protocols \emph{implemented} in a platform; the platform
must either be available as open source or have been described in
sufficient detail in marketing material.  So far all implemented protocols
discussed here assume independence among the failures, selfish behavior,
and subversion of nodes.  This justifies the choice of a \emph{numeric}
trust assumption, expressed only by a fraction of potentially faulty nodes.

It would be readily possible to extend such protocols to more complex fault
assumptions, as formulated by \emph{generic Byzantine quorum
  systems}~\cite{malrei98a}.  For example, this would allow to run
stake-based consensus (as done in some permissionless blockchains) or to
express an arbitrary power structure formulated in a legal agreement for
the consortium~\cite{cachin01}.  No platform offers this yet, however.

\subsection{Hyperledger Fabric -- Apache Kafka and PBFT}
\label{ssec:fabric}

\emph{Hyperledger Fabric} (\url{https://github.com/hyperledger/fabric}) is
a platform for distributed ledger solutions, written in Golang and with a
modular architecture that allows multiple implementations for its
components.  It is one of multiple blockchain frameworks hosted with the
Hyperledger Project (\url{https://www.hyperledger.org/}) and aims at high
degrees of confidentiality, resilience, flexibility, and scalability.

Following ``preview'' releases (\emph{v0.5} and \emph{v0.6}) in 2016, whose
architecture~\cite{cachin16} directly conforms to state-machine
replication, a different and more elaborate design was adopted later and is
currently available in release \emph{v1.0.0-beta}.  The new
architecture~\cite{fabric-ncap16}, termed \emph{Fabric~V1} here, separates
the execution of smart-contract transactions (in the sense of validating
the inputs and outputs of a program) from ordering transactions for
avoiding conflicts (in the sense of an atomic broadcast that ensures
consistency).
This has several advantages, including better scalability, a separation of
trust assumptions for transaction validation and ordering, support for
non-deterministic smart contracts, partitioning of smart-contract code
and data across nodes, and using modular consensus
implementations~\cite{vukoli17}.

The consensus protocol up to release \emph{v0.6-preview} was a native
implementation of PBFT~\cite{caslis02}.  With V1 the \emph{ordering
  service} responsible for conflict-avoidance can be provided by an
\emph{Apache Kafka} cluster (\url{https://kafka.apache.org/}).  Kafka is a
distributed streaming platform with a publish/subscribe interface, aimed at
high throughput and low latency.  It logically consists of \emph{broker
  nodes} and \emph{consistency nodes}, where a set of redundant brokers
processes each message stream and a ZooKeeper instance
(\url{https://zookeeper.apache.org/}) running on the consistency nodes
coordinates the brokers in case of crashes or network problems.  Fabric
therefore inherits is basic resilience against crashes from ZooKeeper.  A
second implementation of the ordering service is under development, which
uses again the PBFT protocol and achieves resilience against subverted
nodes. Besides, BFT-SMaRt (Sec.~\ref{ssec:byzcons}) is currently being
integrated in Fabric~V1 as one of the ordering services.  Since BFT-SMaRt
follows the well-established literature on Byzantine consensus protocols as
mentioned earlier, its properties do not need special discussion here.

\begin{table}[h!]
  \centering
  \begin{tabular}{|l||c|c|c|}
    \hline
    \mbox{} & {\bf Generic} & {\bf Any $t$ nodes} & {\bf Any $f$ nodes}\\
    \mbox{} & {\bf nodes}   & {\bf crash}         & {\bf subverted}\\
    \hline
    {\bf Safety}   & $n$ & $t<n/2$ & \N \\
    {\bf Liveness} & $n$ & $t<n/2$ & \N \\
    \hline
  \end{tabular}
  \caption{Resilience of Hyperledger Fabric~V1 with the Kafka-based orderer.
    It supports an arbitrary number of ``peer'' nodes, but ``node'' refers
    to those nodes comprising the ordering service.  ``Subverted'' means
    an adversarial, Byzantine node.}
  \label{tab:fabrickafka}
\end{table}

\begin{table}[h!]
  \centering
  \begin{tabular}{|l||c|c|c|}
    \hline
    \mbox{} & {\bf Generic} & {\bf Any $t$ nodes} & {\bf Any $f$ nodes}\\
    \mbox{} & {\bf nodes}   & {\bf crash}         & {\bf subverted}\\
    \hline
    {\bf Safety}   & $n$ & $t<n/3$ & $f<n/3$ \\
    {\bf Liveness} & $n$ & $t<n/3$ & $f<n/3$ \\
    \hline
  \end{tabular}
  \caption{Resilience of Hyperledger Fabric v0.6, where ``node'' refers
    to any node.  The same resilience holds for Fabric~V1 with the future
    PBFT-based orderer, where ``node'' refers to those nodes comprising 
    the ordering service.}
  \label{tab:fabricbft}
\end{table}

\subsection{Tendermint}
\label{ssec:tendermint}

\emph{Tendermint Core} (\url{https://github.com/tendermint/tendermint}) is
a BFT protocol that can be best described as a variant of
PBFT~\cite{caslis02}, as its common-case messaging pattern is a variant of
Bracha's Byzantine reliable broadcast~\cite{bracha87}.  In contrast to
PBFT, where the client sends a new transaction directly to all nodes, the
clients in Tendermint disseminate their transactions to the validating
nodes (or, simply, validators) using a gossip protocol.  The external
validity condition, evaluated within the Bracha-broadcast pattern, requires
that a validator receives the transactions by gossip before it can vote for
inclusion of the transaction in a block, much like in PBFT.

Tendermint's most significant departure from PBFT is the continuous
rotation of the leader.  Namely, the leader is changed after every block, a
technique first used in BFT consensus space by the \emph{Spinning} protocol
\cite{scbl09}. Much like Spinning, Tendermint embeds aspects of PBFT's
view-change mechanism into the common-case pattern. This is reflected in
the following: while a validator expects the first message in the Bracha
broadcast pattern from the leader, it also waits for a timeout, which
resembles the view-change timer in PBFT. However, if the timer expires, a
validator continues participating in the Bracha-broadcast message pattern,
but votes for a \emph{nil} block.

Tendermint as originally described by Buchman~\cite{buchma16} suffers from
a livelock bug, pertaining to locking and unlocking votes by validators in
the protocol. However, the protocol contains additional mechanisms not
described in the cited report that prevent the livelock from
occurring~\cite{tendermint17}.  While it appears to be sound, the
Tendermint protocol and its implementation are still subject to a thorough,
peer-reviewed correctness analysis.

\begin{table}[h!]
  \centering
  \begin{tabular}{|l||c|c|c|}
    \hline
    \mbox{} & {\bf Generic} & {\bf Any $t$ nodes} & {\bf Any $f$ nodes}\\
    \mbox{} & {\bf nodes}   & {\bf crash}         & {\bf subverted}\\
    \hline
    {\bf Safety}   & $n$ & $t<n/3$ & $f<n/3$ \\
    {\bf Liveness} & $n$ & $t<n/3$ & $f<n/3$ \\
    \hline
  \end{tabular}
  \caption{Resilience of Tendermint.}
  \label{tab:tendermint}
\end{table}

\subsection{Symbiont -- BFT-SMaRt}
\label{ssec:symbiont}

\emph{Symbiont Assembly} (\url{https://symbiont.io/technology/assembly}) is a
proprietary distributed ledger platform.  The company that stands behind
it, Symbiont, focuses on applications of distributed ledgers in the
financial industry, providing automation for modeling and executing complex
instruments among institutional market participants.

Assembly implements resilient consensus in its platform based on the
open-source BFT-SMaRt toolkit (Sec.~\ref{ssec:byzcons}).
Symbiont uses its own reimplementation of BFT-SMaRt in a different
programming language; it reports performance numbers of 80'000 transactions
per second (tps)
using a 4-node cluster on a LAN.  This matches the throughput expected from
BFT-SMaRt~\cite{besoal14} and similar results in the research literature on
BFT protocols~\cite{agkqv15}.

Assembly uses the standard resilience assumptions for BFT consensus in the
eventually-synchronous model considered here.

\begin{table}[h!]
  \centering
  \begin{tabular}{|l||c|c|c|}
    \hline
    \mbox{} & {\bf Generic} & {\bf Any $t$ nodes} & {\bf Any $f$ nodes}\\
    \mbox{} & {\bf nodes}   & {\bf crash}         & {\bf subverted}\\
    \hline
    {\bf Safety}   & $n$ & $t<n/3$ & $f<n/3$ \\
    {\bf Liveness} & $n$ & $t<n/3$ & $f<n/3$ \\
    \hline
  \end{tabular}
  \caption{Resilience of BFT-SMaRt (reimplemented inside Symbiont Assembly).}
  \label{tab:symbiont-bftsmart}
\end{table}

\subsection{R3 Corda -- Raft and BFT-SMaRt}
\label{ssec:coda}

Unlike most of the other permissioned blockchain platforms discussed here,
\emph{Corda} (\url{https://github.com/corda/corda}) does not order all
transactions as one single virtual execution that forms the blockchain.
Instead, it defines \emph{states} and \emph{transactions}, where every
transaction consumes (multiple) states and produces a new
state~\cite{hearn16}.  Only nodes affected by a transaction store it.  Seen
across all users, this transaction execution model produces a hashed
directed acyclic graph or \emph{Hash-DAG}.  Transactions must be
\emph{valid}, i.e., endorsed by the issuer and other affected nodes and
correct according to the underlying smart-contract logic governing the
state.  Each state points to a \emph{notary} responsible for ensuring
transaction \emph{uniqueness}, i.e., that each state is consumed only once.
The notary is a logical service that can be provided jointly by multiple
nodes.  The \emph{type} of a state may designate an asset represented by
the network, such as a token or an obligation, or anything else controlled
by a smart contract.

A transaction in Corda consumes only states controlled by the same notary;
hence, one notary by itself can atomically verify the transaction's
validity and uniqueness to decide whether it is executed or not.  To enable
transactions that operate across states governed by different notaries,
there is a specialized transaction that changes the notary, such that one
notary will become responsible for validating the transaction.

Since a node stores only a part of the Hash-DAG, it only knows about
transactions and states that concern the node.  This contrasts with most
other distributed ledgers and provides means for partitioning the data
among the nodes.  As is the case for other smart-contract platforms,
transactions refer to contracts that can be programmed in a universal
general-purpose language.

A notary service in Corda orders and timestamps transactions that include
states pointing to it.  ``Notaries are expected to be composed of multiple
mutually distrusting parties who use a standard consensus algorithm''
(\url{https://docs.corda.net}).  A notary service needs to
cryptographically sign its statements of transaction uniqueness, such that
other nodes in the network can rely on its assertions without directly
talking to the notary.
Currently there is support for running a notary service as a single node
(centralized), for running a distributed crash-tolerant implementation
using Raft (Sec.~\ref{ssec:crashcons}), and for distributing it using the
open-source BFT-SMaRt toolkit (Sec.~\ref{ssec:byzcons}).
When using Raft deployed on $n$ nodes, a Corda notary tolerates crashes of
any $t < n/2$ of these nodes (Sec.~\ref{ssec:crashcons}).  With BFT-SMaRt
running on $n$ nodes, the notary is resilient to the subversion of $f <
n/3$ nodes.

\begin{table}[h!]
  \centering
  \begin{tabular}{|l||c|c|c|c|}
    \hline
    \mbox{} & {\bf Notary} & {\bf Any $t$ nodes} & {\bf Any $f$ nodes}\\
    \mbox{} & {\bf nodes}   & {\bf crash}         & {\bf subverted}\\
    \hline
    {\bf Safety}   & $n$ & $t < n/2$ & \N \\
    {\bf Liveness} & $n$ & $t < n/2$ & \N \\
    \hline
  \end{tabular}
  \caption{Resilience of a Raft-based notary service in Corda.
    A Corda network may contain multiple notary services and many 
    more nodes.}
  \label{tab:cordaraft}
\end{table}

\begin{table}[h!]
  \centering
  \begin{tabular}{|l||c|c|c|c|}
    \hline
    \mbox{} & {\bf Notary} & {\bf Any $t$ nodes} & {\bf Any $f$ nodes}\\
    \mbox{} & {\bf nodes}   & {\bf crash}         & {\bf subverted}\\
    \hline
    {\bf Safety}   & $n$ & $t < n/3$ & $f < n/3$ \\
    {\bf Liveness} & $n$ & $t < n/3$ & $f < n/3$ \\
    \hline
  \end{tabular}
  \caption{Resilience of a BFT-SMaRt-based notary service in Corda.
    A Corda network may contain multiple notary services and many 
    more nodes.}
  \label{tab:cordabft}
\end{table}

\subsection{Iroha -- Sumeragi}
\label{ssec:iroha}

\emph{Iroha} (\url{https://github.com/hyperledger/iroha}) is another open-source
blockchain platform developed under the Hyperledger Project.  Its architecture
is inspired by the original~(v0.6) design of Fabric (Sec.~\ref{ssec:fabric}).
All validating nodes collaboratively execute a Byzantine consensus protocol.
In that sense it is also similar to Tendermint and Symbiont Assembly.

The \emph{Sumeragi} consensus library of Iroha is ``heavily inspired'' by
BChain~\cite{dmpz14}
a chain-style Byzantine replication protocol that propagates transactions
among the nodes with a ``chain'' topology.  Chain
replication~\cite{rensch04,CharronBostPS10} arranges the $n$ nodes linearly
and each node normally only receives messages from its predecessor and
sends messages to its successor.  Although there is a leader at the head of
the chain, like in many other protocols, the leader does not become a
bottleneck since it usually communicates only with the head and the tail of
the chain, but not with all $n$~nodes.  This balances the load among the
nodes and lets chain-replication protocols achieve the best possible
throughput~\cite{glpq10,agkqv15}, at the cost of higher normal-case latency
and slightly increased time for reconfiguration after faults.

In Sumeragi, the order of the nodes is determined based on a reputation
system, which takes the ``age'' of a node and its past performance into
account.

As becomes apparent from the online documentation
(\url{https://github.com/hyperledger/iroha/wiki/Sumeragi}),
though, the protocol departs from the ``chain'' pattern, because the leader
``broadcasts'' to all nodes and so does the node at the tail.  Hence, it is
neither BChain nor chain replication.  Assuming that Sumeragi would
correctly implement BChain, then it relies on the standard assumptions for
BFT consensus in the eventually-synchronous model, just like Fabric~v0.6,
Tendermint, and Symbiont.

\begin{table}[h!]
  \centering
  \begin{tabular}{|l||c|c|c|}
    \hline
    \mbox{} & {\bf Generic} & {\bf Any $t$ nodes} & {\bf Any $f$ nodes}\\
    \mbox{} & {\bf nodes}   & {\bf crash}         & {\bf subverted}\\
    \hline
    {\bf Safety}   & $n$ & $t<n/3$ & $f<n/3$ \\
    {\bf Liveness} & $n$ & $t<n/3$ & $f<n/3$ \\
    \hline
  \end{tabular}
  \caption{Resilience of Iroha, assuming the Sumeragi consensus
    implementation is BChain~\cite{dmpz14}.}
  \label{tab:iroha}
\end{table}

\subsection{Kadena -- Juno and ScalableBFT}
\label{ssec:kadena}

\emph{Juno} from \emph{kadena} (\url{https://github.com/kadena-io/juno}) is a platform for running
smart contracts that has been developed until about November 2016 according
to its website.  Juno claims to use a ``Byzantine Fault Tolerant Raft''
protocol for consensus and appears to address the standard BFT model with
$n$ nodes, $f < n/3$ Byzantine faults among them, and eventual
synchrony~\cite{dwlyst88} as timing assumption.
Later Juno has been deprecated in favor of a ``proprietary BFT-consensus
protocol'' called \emph{ScalableBFT}~\cite{martin16}, which is ``inspired
by the Tangaroa protocol'' and optimizes performance compared to Juno and
Tangaroa.  The whitepaper cites over 7000 transactions per second (tps)
throughput on a cluster with size 256 nodes.

The design and implementation of ScalableBFT are proprietary and not
available for public review.  Being based on Tangaroa, the design might
suffer from its devastating problems mentioned in
Section~\ref{ssec:byzcons}.  Further statements about ScalableBFT made in a
blog post~\cite{samman16} do not enhance the trust in its safety: ``Every
transaction is replicated to every node. When a majority of nodes have
replicated the transaction, the transaction is committed.''  As is
well-known from the literature~\cite{CharronBostPS10,CachinGR11} in the
model considered here, with public-key cryptography for message
authentication and asynchrony, agreement in a consensus protocol can only
be ensured with $n > 3f$ and Byzantine quorums~\cite{malrei98a} of size
\emph{strictly larger} than $\frac{n+f}{2}$, which reduces to $2f+1$ with
$n = 3f+1$ nodes.  Hence ``replicating among a majority'' does \emph{not}
suffice.

The claimed performance number of more than 7000 tps is in line with the
throughput of 30'000--80'000 tps, as reported by a representative
state-of-the-art BFT protocol evaluation in the literature~\cite{agkqv15}.
However, since Juno is proprietary, it is not not clear how it actually
works nor why one should trust it, as discussed before.  One should rather
build on established consensus approaches and publicly validated algorithms
than on a proprietary protocol for resilience.

As the resilience of Juno and ScalableBFT cannot be assessed, and as it
remains unclear whether it actually provides consensus as intended, there
is no summary table.

\subsection{Chain -- Federated Consensus}
\label{ssec:chain}

The \emph{Chain Core} platform~(\url{https://chain.com}) is a generic
infrastructure for an institutional consortium to issue and transfer
financial assets on permissioned blockchain networks.  It focuses on the
financial services industry and supports multiple different assets within
the same network.

The \emph{Federated Consensus}~\cite{chainwp17} protocol of Chain Core is
executed by the $n$ nodes that make up the network.  One of the nodes is
statically configured as ``block generator.''  It periodically takes a
number of new, non-executed transactions, assembles them into blocks, and
submits the block for approval to ``block signers.''  Every signer
validates the block proposed for a given block height, checking the
signature of the generator, validating the transactions, and verifying some
real-time constraints and then signs an endorsement for the block.  Each
signer endorses only one block at each height.  Once a node receives $q$
such endorsements for a block, the node appends the block to its chain.

The protocol is resilient to a number of malicious (Byzantine-faulty)
signers but not to a malicious block generator.  If the block generator
violates the protocol (e.g., by signing two different blocks for the same
block height) the ledger might fork (i.e., the consensus protocol violates
safety).  The documentation states that such misbehavior should be
addressed by retaliation and measures for this remain outside the protocol.

More specifically, when assuming the block generator operates correctly and
is live, this Federated Consensus reduces to an ordinary Byzantine quorum
system that tolerates $f$ faulty signer nodes when $q = 2f+1$ and $n =
3f+1$; its use for consensus is similar, say, to the well-understood
``authenticated echo broadcast''~\cite[Sec.~3.10.3]{CachinGR11}.  Up to $f$
block signers may behave arbitrarily, such as by endorsing incorrect
transactions or by refusing to participate, and the protocol will remain
live and available (with the correct block generator).

Overall, however, Federated Consensus is a special case of a standard
BFT-consensus protocol that appears to operate with a fixed ``leader'' (in
the role of the block generator).  The protocol \emph{cannot} prevent forks
if the generator is malicious.  Even if the generator simply crashes, the
protocol halts and requires manual intervention.  Standard BFT protocols
instead will tolerate leader corruption and automatically switch to a
different leader if it becomes apparent that one leader malfunctions.

Since the block generator must be correct, the purpose of a signature
issued by a block signer remains unclear, at least at the level of the
consensus protocol.  The only reason appears to be guaranteeing that the
signer cannot later repudiate having observed a block.

\begin{table}[h!]
  \centering
  \begin{tabular}{|l||c|c|c||c|c|c|}
    \hline
    \mbox{} & {\bf Generic} & {\bf Any $t$ nodes} & {\bf Any $f$ nodes} &
              {\bf Special} & {\bf Any $s$ special} & {\bf Special nodes} \\
    \mbox{} & {\bf nodes} & {\bf crash} & {\bf subverted} &
              {\bf nodes} & {\bf nodes crash} & {\bf subverted} \\
    \hline
    {\bf Safety}   & $n$ & $t<n/3$ & $f < n/3$ & 
                     $m=1$ & \N & \N \\
    {\bf Liveness} & $n$ & $t<n/3$ & $f < n/3$ & 
                     $m=1$ & \N & \N \\
    \hline
  \end{tabular}
  \caption{Resilience of Federated Consensus in Chain Core.
    The single special node is the block generator.  The protocol cannot
    handle the case that it fails.}
  \label{tab:chaincore}
\end{table}

\subsection{Quorum -- QuorumChain and Raft}
\label{ssec:quorum}

\emph{Quorum} (\url{https://github.com/jpmorganchase/quorum}), mainly from
developers at JPMorgan Chase, is an enterprise-focused version of Ethereum,
executing smart contracts with the Ethereum virtual machine, but using an
alternative to the default proof-of-work consensus protocol of the public
Ethereum blockchain.  The platform currently contains two consensus
protocols, called \emph{QuorumChain} and \emph{Raft-based consensus}.

\paragraph{QuorumChain.}
This protocol uses a smart contract to validate blocks.  The trust model
specifies a set of $n$ ``voter'' nodes and some number of ``block-maker''
nodes, whose identities are known to all nodes.  The documentation remains
unclear about the trust model, not clearly expressing in which ways one or
more of these nodes might fail or behave adversarially.  (One can draw some
conclusions from the protocol though.)

The protocol uses the standard peer-to-peer gossip layer of Ethereum to
propagate blocks and votes on blocks, but the logic itself is formulated as
a smart contract deployed with the genesis block.  Nodes digitally sign
every message they send.  Only block-maker nodes are permitted to propose
block to be appended; nodes with voter role validate blocks and express
their approval by a (yes) vote.  A block-maker waits for a randomly chosen
time and then creates, signs, and propagates a new block that extends its
own chain.  A voter will validate the block (by executing its transactions
and checking its consistency), ``vote'' on it, and propagate this.
A voter apparently votes for every received block that is valid and extends
its own chain, and it may vote multiple times for a given block height.
Voting continues for a period specified in real time.  Each node accepts
and extends its own chain with the block that obtains more votes than a
given threshold, and if there are multiple ones, the one with most votes.
There is one block-maker node by default.

To assess the resilience of the protocol, it is obvious that already one
malicious block-maker node can easily create inconsistencies (chain forks)
unless the network is perfect and already provides consensus.  With one
block-maker, if this node crashes, the protocol halts.  Depending on how
the operator sets the voting threshold and on the network connectivity, it
may fork the chain with only two block-makers and without any Byzantine
fault.  With a Byzantine fault in a block-maker node or a voter node can
disrupt the protocol and also create inconsistencies.  Furthermore, the
protocol relies on synchronized clocks for safety and liveness.  Taken
together, the protocol cannot ensure consensus in any realistic sense.

\begin{table}[h]
  \centering
  \begin{tabular}{|l||c|c|c||c|c|c|}
    \hline
    \mbox{} & {\bf Generic} & {\bf Any $t$ nodes} & {\bf Any $f$ nodes} &
              {\bf Special} & {\bf Any $s$ special} & {\bf Special nodes} \\
    \mbox{} & {\bf nodes} & {\bf crash} & {\bf subverted} &
              {\bf nodes} & {\bf nodes crash} & {\bf subverted} \\
    \hline
    {\bf Safety}   & $n$ & $t<n/3$ & $f < n/3$ & 
                     $m=1$ & \N & \N \\
    {\bf Liveness} & $n$ & $t<n/3$ & $f < n/3$ & 
                     $m=1$ & \N & \N \\
    \hline
  \end{tabular}
  \caption{Resilience of QuorumChain consensus in Quorum.
    The special nodes are the block-maker nodes.
    With $m>1$ block-maker nodes, safety is not guaranteed and 
    forks may occur due to network effects, even when all block-makers
    are correct.}
  \label{tab:quorumchain}
\end{table}

\paragraph{Raft-based consensus.}
The second and more recent consensus option available for Quorum is based
on the Raft protocol~\cite{ongous14}, which is a popular variant of
Paxos~\cite{lampor98} available in many open-source toolkits.  Quorum uses
the implementation in \emph{etcd} (\url{https://github.com/coreos/etcd})
and co-locates every Quorum-node with an etcd-node (itself running Raft).
Raft will replicate the transactions to all participating nodes and ensure
that each node locally outputs the same sequence of transactions, despite
crashes of nodes.  The deployment actually tolerates that any $t < n/2$ of
the $n$ etcd-nodes may crash.  Raft relies on timeliness and synchrony only
for liveness, not for safety.

This is a canonical design, directly interpreting the replication of Quorum
smart contracts as a replicated state machine.  It seems appropriate for a
protected environment, which is not subject to adversarial nodes.

\begin{table}[h]
  \centering
  \begin{tabular}{|l||c|c|c|c|}
    \hline
    \mbox{} & {\bf Generic} & {\bf Any $t$ nodes} & {\bf Any $f$ nodes}\\
    \mbox{} & {\bf nodes}   & {\bf crash}         & {\bf subverted}\\
    \hline
    {\bf Safety}   & $n$ & $t < n/2$ & \N \\
    {\bf Liveness} & $n$ & $t < n/2$ & \N \\
    \hline
  \end{tabular}
  \caption{Resilience of Raft-based consensus in Quorum.  Every generic
    node is also an etcd-node.}
  \label{tab:quorumraft}
\end{table}

\subsection{MultiChain}
\label{ssec:multichain}

The \emph{MultiChain} platform (\url{https://github.com/MultiChain/multichain}) is
intended for permissioned blockchains in the financial industry and for
multi-currency exchanges in a consortium, aiming at compatibility with the
Bitcoin ecosystem as much as possible.

MultiChain uses a dynamic permissioned model~\cite{greens16}: There is a
list of permitted nodes in the network at all times, identified by their
public keys.  The list can be changed through transactions executed on the
blockchain, but at all times, only nodes on this list validate blocks and
participate in the protocol.

As the MultiChain platform is derived from Bitcoin, its consensus mechanism
is called ``mining''~\cite{greens16}; however, in the permissioned model,
the nodes do not solve computational puzzles.  Instead, any permitted node
may generate new blocks after waiting for a random timeout, subject to a
\emph{diversity} parameter $\rho \in [0,1]$ that constrains the acceptable
miners for a given block height.  More precisely, if the permitted list has
length~$L$, then a block proposal from a node is only accepted if the
blockchain held by the validating node does not already contain a block
generated by the \emph{same} node among the $\lceil \rho L \rceil$ most
recent blocks.  Any participating node will extend its blockchain with the
first valid block of this kind that it receives, and if it learns about
different, conflicting chain extensions, it will select the longer one (as
in Bitcoin).  Furthermore, a well-behaved node will not generate a new
block if its own chain already contains a block of his within the last
$\lceil \rho L \rceil$ blocks.

It appears that the random timeouts and network uncertainty easily lead to
forks in the ledger, even if all nodes are correct.  If two different nodes
may generate a valid block at roughly the same time, and any other node
will append the one of which it hears first to its chain, then these two
nodes will be forked.  This is not different from consensus in Bitcoin and
will eventually converge to a single chain if all nodes follow the
protocol.  However, if a single attacking node generates transactions and
blocks as it wants, and assuming that the network behaves favorably for the
attack, the node can take over the entire network and revert arbitrarily
many past transactions (in the same way as a ``51\%-attack'' in Bitcoin).

Hence, MultiChain exhibits non-final transactions similar to any
proof-of-work consensus.  But whereas lack of finality appears to be a
consequence of the public nature of proof-of-work, and since MultiChain is
permissioned, forks and non-final decisions could be avoided here
completely. The traditional consensus protocols for this model, discussed
in Sections~\ref{ssec:crashcons} and \ref{ssec:byzcons}, all reach
consensus with finality.  In the model of non-final consensus decisions,
with the corresponding delays and throughput constraints, the MultiChain
consensus protocol can only remain consistent and live with one single
correct node.

\begin{table}[h]
  \centering
  \begin{tabular}{|l||c|c|c|c|}
    \hline
    \mbox{} & {\bf Generic} & {\bf Any $t$ nodes} & {\bf Any $f$ nodes}\\
    \mbox{} & {\bf nodes}   & {\bf crash}         & {\bf subverted}\\
    \hline
    {\bf Safety*}   & $n$ & $t < n$ & \N \\
    {\bf Liveness} & $n$ & $t < n$ & \N \\
    \hline
  \end{tabular}
  \caption{Resilience of MultiChain consensus.
    \emph{Safety*} denotes the non-final consistency notion, as achieved by
    proof-of-work consensus, and should be understood formally as in recent
    work on the subject~\cite{gakile15}.
  }
  \label{tab:multichain}
\end{table}

\subsection{Further platforms}

Another extension of the Ethereum platform is \emph{HydraChain}
(\url{https://github.com/HydraChain/hydrachain/blob/develop/README.md}),
which adds support for creating a permissioned distributed ledger using the
Ethereum infrastructure.  The repository describes a proprietary consensus
protocol ``initially inspired by Tendermint.''
Without clear explanation of the protocol and formal review of its
properties, its correctness remains unclear.

The \emph{Swirlds hashgraph algorithm} is built into a proprietary
``distributed consensus platform'' (\url{https://www.swirlds.com}); a white
paper is available~\cite{baird16} and the protocol is also implemented in
an open-source consensus platform for distributed applications, called
\emph{Babble}~(\url{https://github.com/babbleio/babble}).  It targets
consensus for a permissioned blockchain with $n$ nodes and $f < n/3$
Byzantine faults among them, i.e., the standard Byzantine consensus problem
according to Section~\ref{ssec:byzcons}.  In contrast to PBFT and other
protocols discussed there, it operates in a ``completely asynchronous''
model.  The white paper states arguments for the safety and liveness of the
protocol and explains that hashgraph consensus is randomized to circumvent
the FLP impossibility~\cite{filypa85}.  Since the algorithm is guaranteed
to reach agreement on a binary decision (i.e., with only 0/1 outcomes) only
with exponentially small probability in~$n$~\cite[Thm.~5.16]{baird16}, it
appears similar to Ben-Or-style randomized
agreement~\cite[Sec.~5.5]{CachinGR11}.
However, no independent validation or analysis of hashgraph consensus is
available.

\section{Permissionless blockchains}
\label{sec:public}

The most widely used consensus protocols in permissionless blockchains are
\emph{proof-of-work} and \emph{proof-of-stake}, which appear always coupled
to a cryptocurrency.  There is a lot of research and development activity
addressing them at the moment, and covering this would go beyond the scope
of this text.

Instead, this section reviews some notable variants of protocols that do
not rely on a strict notion of membership and therefore differ from the
permissioned blockchain platforms reviewed in the last section.  The
protocols described here depart in other ways from the traditional
consensus notions (crash-tolerant, Byzantine, and proof-of-work-style
consensus).  Conceptually they fall somewhere between the extremes of a BFT
protocol and Nakamoto's proof-of-work consensus.

\subsection{Sawtooth Lake -- Proof of Elapsed Time}

The Hyperledger Sawtooth
platform~(\url{https://github.com/hyperledger/sawtooth-core}) provides
means for running general-purpose smart contracts on a distributed ledger.
It can use a permissioned and a public, permissionless mode.  The platform
also introduces a novel consensus protocol called \emph{Proof of Elapsed
  Time (PoET)}, originally contributed by Intel, which is based on the
insight that proof-of-work essentially imposes a mandatory but random
\emph{waiting time} for leader election.

In particular, when ignoring the mining reward of Bitcoin, Nakamoto
consensus lets all nodes participate in a probabilistic experiment, where
each node is delayed for a random duration.  Once the timer expires, the
node can \emph{prove} to all others in a verifiable way that it has
executed the ``waiting step'' correctly for extending the blockchain.  The
node propagates its solution to all others as quickly as possible, because
only the longest chain is valid.  With the correct relation between the
tunable waiting-delay and the expected time for reaching every node with
the new solution, this creates a stable consensus protocol.  The Bitcoin
network's operation and mathematical analysis~\cite{gakile15} demonstrate
this.

PoET consensus executes the waiting step in a trusted hardware module, the
\emph{Intel Software Guard Extensions (SGX)} available in many modern Intel
CPUs.  Every node essentially calls an \emph{enclave} inside SGX for
generating a random delay, waiting accordingly, and then declaring itself
to be the leader in consensus and extending the blockchain.  The platform
creates an attestation that can be used by any node to verify that the
leader correctly waited for the proper random time.  Assuming the hardware
module cannot be subverted, this creates the same kind of non-final
consensus as with mining in proof-of-work.  

The energy waste caused by mining goes away.  However, economic investment
still increases the influence on the protocol because the probability of a
node becoming the leader is proportional to the number of hardware modules
under its control.  PoET is compatible with permissionless blockchains, but
only assuming an unlimited supply of trusted modules.  As the protocol's
security depends on the hardware module potentially running on an
adversarial host, the impact of attacks will have to be understood as well
(e.g., SGX is susceptible to rollback attacks~\cite{bclk17} and key
extraction~\cite{swgmm17}).

In a permissioned setting, the participating modules could be authenticated
and the weight of a node can be fixed through this.  However, with known
nodes, traditional BFT consensus protocols have several advantages compared
to PoET: they are more efficient, do not rely on a single vendor's
hardware, and create final decisions.  Moreover, if trusted modules are
available, then BFT consensus can increase the resilience to $f < n/2$
subverted nodes and achieve more than 70'000 tps throughput in a
LAN~\cite{kbcdkm12}.

\subsection{Ripple and Stellar}

Ripple~(\url{https://ripple.com/}) and
Stellar~(\url{https://www.stellar.org/}) are two globally operating
exchange networks with built-in cryptocurrencies; unlike Bitcoin, they do
not involve mining and operate in a somewhat permissioned fashion.  The
\emph{Ripple protocol consensus algorithm (RPCA)} and its offspring
\emph{Stellar consensus protocol (SCP)} depart from the traditional
security assumption for consensus protocols (i.e., some $f < n/3$ faulty
nodes) by making their trust assumptions
\emph{flexible}~\cite{scyobr14,mazier16}.  This means that each node would
declare on its own which nodes it trusts, instead of accepting a global
assumption on which node collusions the protocol tolerates.  Each node
designates a list of other nodes sufficient to \emph{convince} itself
(through the \emph{unique node list} in Ripple or the \emph{quorum slice}
of Stellar).

Ripple and Stellar each maintain one distributed ledger governed by the
protocol, which records exchanges on the respective network.

\paragraph{Ripple.}
In Ripple, the process of advancing the common distributed ledger is
controlled by so-called validating nodes.  They periodically start to
create a new ledger entry (every few seconds) and iteratively vote in
rounds on its content; each node accepts a proposed ledger update if 50\%,
\dots, 80\% (increasing by $+10\%$ per round) of the signed updates that it
receives match.

Ripple's documentation states that $4/5 \cdot n$ of \emph{all} $n$
validator nodes must be correct for maintaining
correctness~\cite{scyobr14}.
This would correspond to tolerating $f < n/5$ subverted nodes in
traditional BFT systems.

Furthermore, it is obvious that a minimal overlap among the
\emph{convincing-sets} (i.e., unique node lists) of all pairs of validating
nodes is required, since otherwise they could exhibit split-brain behavior
and the ledger would fork.  Ripple states that the overlap should be at
least $1/5$ of the size of the larger list~\cite{scyobr14}.
The only peer-reviewed analysis of the Ripple protocol, however,
contradicts this.  Armknecht et al.~\cite{akmyz15} show that when each node
has $\rho n$ validators in its convincing-set, then ledger forks are ruled
out only if for every two nodes, their lists contain \emph{more than} $2
(1-\rho) m$ common nodes, where $m$ is the size of the larger of the two
lists.  This implies more than $2/5$ overlap with $\rho = 0.8$ as chosen by
the Ripple network.

Currently, Ripple ``provides a default and recommended list of validators
operated by Ripple and third parties,'' through a static configuration
file;
by default there are \emph{five} validators operated by Ripple which trust
each other and no other node.  (``At present, Ripple cannot recommend any
validators aside from the 5 core validators run by Ripple (the
company)''~\cite{ripple17}.)
It appears that this list is adopted by most validating nodes in the
system; consequently, trust is by far not as decentralized as advertised.

A typical consensus process creating one new ledger entry completes in less
than four seconds on average.  Ripple has stated a throughput of about 1000
tps on a test network~\cite{ripple17blog}.
Compared to several 10'000 tps achievable on traditional BFT platforms with
a small group of 4--10 validators~\cite{soubes15}, this seems unnecessarily
slow.

\paragraph{Stellar.}
As Stellar evolved from Ripple, it uses similar ideas and a protocol 
called \emph{federated Byzantine agreement}~\cite{mazier16} within the
Stellar consensus protocol (SCP).  Only \emph{validator nodes} participate
in the protocol for reaching consensus.  Each validator declares its own
\emph{convincing-set} (called ``quorum slice'' and similar to Ripple's
unique node list) that must sufficiently overlap with the convincing-sets
of other nodes for preventing forks.  A node accepts a ``vote'' or a
transaction for the ledger when a threshold of nodes in its convincing-set
confirm it.
Examples in the documentation and the white paper~\cite[Fig.~3]{mazier16}
suggest the use of hierarchical structures with different groups organized
into multiple levels, where a different threshold may exist for each group
(but the ``threshold should be 2/3 for the top level'').

For instance, a convincing-set (i.e., a quorum slice) in a hierarchy with
two levels could be like this (expressed in percent and rounded to
integers):

\quad
\scalebox{0.8}{\begin{minipage}{8cm}
\begin{align*}
  \text{67\% of}\ & \text{\var{Groups}, where}\\
  & \var{Groups} = \{\var{Banks}, \var{Auditors}, \var{Advisors}, 
    \var{Friends}\} &
  \qquad\qquad\text{// here: 3 of 4 sub-groups} \\
  \text{51\% of}\ & \text{\var{Banks}, where}\\
  & \var{Banks} = \{\var{Bank-1}, \var{Bank-2}, \var{Bank-3} \}
  &\text{// here: 2 of 3 banks} \\
  \text{58\% of}\ & \text{\var{Auditors}, where}\\
  & \var{Auditors} = \{A, B, C, D, E, F, G\} 
  &\text{// here: 5 of 7 auditors} \\
  \text{51\% of}\ & \text{\var{Advisors}, where}\\
  & \var{Advisors} = \{\var{1}, \var{2}, \var{3} \}
  &\text{// here: 2 of 3 advisors} \\
  \text{ 1\% of}\ & \text{\var{Friends}, where}\\
  & \var{Friends} = \{\var{Alice}, \var{Bob}, \var{Charlie}, \dots, 
    \var{Zach} \}
  &\text{// here: 1 of 26 friends}\\
\end{align*}
\end{minipage}}

Similar structures have been known as \emph{Byzantine quorum systems
  (BQS)}~\cite{malrei98a} and are well-understood.  They can readily be
used to build consensus protocols for BFT systems~\cite{cachin01}.  The
documentation available for Stellar does not relate to this literature,
however.

Furthermore, it seems that for constructing one single ledger, the
convincing-sets for all useful configurations of SCP should intersect at
the top of the suggested hierarchies.  This appears to introduce some
amount of centralization, similar to BQS~\cite{marewo00}.

At this time, determining the similarities and differences
between the quorum slices of SCP and generic BQS for Byzantine consensus
remains an open problem.

\subsection{IOTA Tangle}

IOTA (\url{http://iota.org/}) is heralded as a ``cryptocurrency without a
blockchain'' and creates a Hash-DAG instead, which is called the
\emph{tangle}~\cite{popov16}.  All of its tokens are created at the outset.
Transactions are propagated in a peer-to-peer network like in Bitcoin; each
transaction transfers tokens owned by a node to others and must be signed
by the owner's key node.  A transaction also includes the solution to a
proof-of-work puzzle and \emph{approves} two or any $k \geq 2$ transactions
by including a hash of them in the transaction.  This creates the DAG, with
an edge pointing \emph{from} each confirmed transaction \emph{to} the new
one.  A weight is assigned to the transaction proportionally to the
difficulty of the puzzle that the node has solved for producing it.  The
node is supposed to choose the $k$ transactions to confirm randomly, from
all transactions it is aware of, and to verify the two transactions plus
\emph{all} transitive predecessor transactions of them.
Implicitly it should also verify the transaction(s) that have earlier
assigned the tokens to the node.

A node can cheat by (1) issuing an invalid transaction (double-spending),
(2) including invalid predecessor transactions, or (3) not selecting the
transactions to confirm randomly.  However, other nodes would intuitively
not include invalid transactions produced from (1) or (2), and therefore
such a transaction would become orphaned.  In a dense graph, it is expected
that most transactions of a certain age are approved by a vast majority of
all newly generated transactions.

How strongly a new transaction approves its predecessor transactions
depends on the weight, i.e., the computational work, that went into
producing it.  The stability of the system therefore rests on the
assumption that the majority of the computing power among the nodes behaves
correctly.  Furthermore, if there are conflicting parts of the graph, each
node will decide on its own which side to trust.  This is done through a
probabilistic sampling algorithm and deciding according to a statistical
test.  The construction of the DAG is reminiscent of Lewenberg et
al.'s~\cite{lesozo15} \emph{inclusive} blockchain protocols, but the
details differ.

The white paper~\cite{popov16} and documentation claim that the tangle
hash-DAG ensures a similar level of consistency as other permissionless
blockchain systems.  No publicly reviewed, formal analysis is available,
however.  Without any independent assessment of the protocol's properties,
it remains unclear how strictly the tangle emulates a notion of consensus
among the nodes.

\section{Conclusion}

This paper has summarized some of the most prominent blockchain consensus
protocols, focusing on permissioned systems in the sense that their
participants are identified.

We have argued that developing consensus protocols is similar to
engineering cryptographic systems, and that blockchain developers should
look towards the established experience in cryptography, security,
and the theory of distributed systems for
building trustworthy systems.  Otherwise, it might be dangerous to entrust
financial value to new protocols.  Open discussion, expert reviews, broad
validation, and standards recommendations should be employed.

The overview of consensus protocols and their properties contributes to
this effort, by establishing a common ground for formal protocol reviews
and more technical comparisons.  Once sufficiently many systems become
available publicly and are widely used, it will be interesting to compare
their performance through benchmarks and to observe their resilience to
actual attacks or network incidents.

\section*{Acknowledgments}

We thank Andreas Kind, Pedro Moreno-Sanchez, and Bj\"{o}rn Tackmann for
interesting discussions and valuable comments.

This work has been supported in part by the European Commission through the
Horizon 2020 Framework Programme (H2020-ICT-2014-1) under grant agreements
number 643964-SUPERCLOUD and 644579~ESCUDO-CLOUD and in part by the Swiss
State Secretariat for Education, Research and Innovation (SERI) under
contracts number 15.0091 and~15.0087.


\begin{thebibliography}{10}

\bibitem{ripple17blog}
W.~Anderson.
\newblock Ripple consensus ledger can sustain 1000 transactions per second.
\newblock Ripple Dev Blog,
  \url{https://ripple.com/dev-blog/ripple-consensus-ledger-can-sustain-1000-transactions-per-second/},
  2017.

\bibitem{fabric-ncap16}
E.~Androulaki, C.~Cachin, K.~Christidis, C.~Murthy, B.~Nguyen, and
  M.~Vukoli{\'c}.
\newblock Next consensus architecture proposal.
\newblock Hyperledger Wiki, Fabric Design Documents, available at
  \url{https://github.com/hyperledger/fabric/blob/master/proposals/r1/Next-Consensus-Architecture-Proposal.md},
  2016.

\bibitem{akmyz15}
F.~Armknecht, G.~O. Karame, A.~Mandal, F.~Youssef, and E.~Zenner.
\newblock Ripple: Overview and outlook.
\newblock In M.~Conti, M.~Schunter, and I.~G. Askoxylakis, editors, {\em Proc.\
  Trust and Trustworthy Computing (TRUST)}, volume 9229 of {\em Lecture Notes
  in Computer Science}, pages 163--180. Springer, 2015.

\bibitem{AttiyaW04}
H.~Attiya and J.~Welch.
\newblock {\em Distributed Computing: Fundamentals, Simulations and Advanced
  Topics}.
\newblock Wiley, second edition, 2004.

\bibitem{agkqv15}
P.~Aublin, R.~Guerraoui, N.~Knezevic, V.~Qu{\'{e}}ma, and M.~Vukolic.
\newblock The next 700 {BFT} protocols.
\newblock {\em ACM Transactions on Computer Systems}, 32(4):12:1--12:45, 2015.

\bibitem{alrl04}
A.~Avizienis, J.-C. Laprie, B.~Randell, and C.~Landwehr.
\newblock Basic concepts and taxonomy of dependable and secure computing.
\newblock {\em IEEE Transactions on Dependable and Secure Computing},
  1(1):11--33, 2004.

\bibitem{baird16}
L.~Baird.
\newblock The {Swirlds} hashgraph consensus algorithm: Fair, fast, {Byzantine}
  fault tolerance.
\newblock Swirlds Tech Report SWIRLDS-TR-2016-01, available online,
  \url{http://www.swirlds.com/developer-resources/whitepapers/}, 2016.

\bibitem{bessou12}
A.~Bessani and J.~Sousa.
\newblock From {Byzantine} consensus to {BFT} state machine replication: A
  latency-optimal transformation.
\newblock In {\em Proc.\ 9th European Dependable Computing Conference}, pages
  37--48, 2012.

\bibitem{besoal14}
A.~N. Bessani, J.~Sousa, and E.~A.~P. Alchieri.
\newblock State machine replication for the masses with {BFT-SMaRt}.
\newblock In {\em Proc.\ 44th International Conference on Dependable Systems
  and Networks}, pages 355--362, 2014.

\bibitem{JonesPM16}
B.~Beyer, C.~Jones, J.~Petoff, and N.~R. Murphy, editors.
\newblock {\em Site Reliability Engineering: How Google Runs Production
  Systems}.
\newblock O'Reilly, Sebastopol, 2016.

\bibitem{bracha87}
G.~Bracha.
\newblock Asynchronous {Byzantine} agreement protocols.
\newblock {\em Information and Computation}, 75:130--143, 1987.

\bibitem{bclk17}
M.~Brandenburger, C.~Cachin, M.~Lorenz, and R.~Kapitza.
\newblock Rollback and forking detection for trusted execution environments
  using lightweight collective memory.
\newblock e-print, arXiv:1701.00981 [cs.DC], 2017.

\bibitem{buchma16}
E.~Buchman.
\newblock Tendermint: {Byzantine} fault tolerance in the age of blockchains.
\newblock M.Sc.~Thesis, University of Guelph, Canada, June 2016.

\bibitem{tendermint17}
E.~Buchman and J.~Kwon.
\newblock Private discussion, 2017.

\bibitem{cachin01}
C.~Cachin.
\newblock Distributing trust on the {Internet}.
\newblock In {\em Proc.\ International Conference on Dependable Systems and
  Networks (DSN-DCCS)}, pages 183--192, 2001.

\bibitem{cachin09}
C.~Cachin.
\newblock Yet another visit to {Paxos}.
\newblock Research Report RZ 3754, IBM Research, Nov. 2009.

\bibitem{cachin16}
C.~Cachin.
\newblock Architecture of the {Hyperledger} blockchain fabric.
\newblock Workshop on Distributed Cryptocurrencies and Consensus Ledgers (DCCL
  2016), 2016.
\newblock URL: \url{https://www.zurich.ibm.com/dccl/papers/cachin_dccl.pdf}.

\bibitem{CachinGR11}
C.~Cachin, R.~Guerraoui, and L.~Rodrigues.
\newblock {\em Introduction to Reliable and Secure Distributed Programming
  ({Second Edition})}.
\newblock Springer, 2011.

\bibitem{ckps01}
C.~Cachin, K.~Kursawe, F.~Petzold, and V.~Shoup.
\newblock Secure and efficient asynchronous broadcast protocols (extended
  abstract).
\newblock In J.~Kilian, editor, {\em Advances in Cryptology:\ CRYPTO 2001},
  volume 2139 of {\em Lecture Notes in Computer Science}, pages 524--541.
  Springer, 2001.

\bibitem{cakush05}
C.~Cachin, K.~Kursawe, and V.~Shoup.
\newblock Random oracles in {Constantinople}: Practical asynchronous
  {Byzantine} agreement using cryptography.
\newblock {\em Journal of Cryptology}, 18(3):219--246, 2005.

\bibitem{cacpor02}
C.~Cachin and J.~A. Poritz.
\newblock Secure intrusion-tolerant replication on the {Internet}.
\newblock In {\em Proc.\ International Conference on Dependable Systems and
  Networks (DSN-DCCS)}, pages 167--176, June 2002.

\bibitem{caslis02}
M.~Castro and B.~Liskov.
\newblock Practical {B}yzantine fault tolerance and proactive recovery.
\newblock {\em ACM Transactions on Computer Systems}, 20(4):398--461, Nov.
  2002.

\bibitem{chainwp17}
Chain protocol whitepaper.
\newblock Available online,
  \url{https://chain.com/docs/1.2/protocol/papers/whitepaper}, 2017.

\bibitem{chgrre07}
T.~D. Chandra, R.~Griesemer, and J.~Redstone.
\newblock {Paxos} made live: An engineering perspective.
\newblock In {\em Proc.\ 26th ACM Symposium on Principles of Distributed
  Computing (PODC)}, pages 398--407, 2007.

\bibitem{CharronBostPS10}
B.~Charron-Bost, F.~Pedone, and A.~Schiper, editors.
\newblock {\em Replication: Theory and Practice}, volume 5959 of {\em Lecture
  Notes in Computer Science}.
\newblock Springer, 2010.

\bibitem{chkevi01}
G.~V. Chockler, I.~Keidar, and R.~Vitenberg.
\newblock Group communication specifications: A comprehensive study.
\newblock {\em ACM Computing Surveys}, 33(4):427--469, 2001.

\bibitem{cwadm09}
A.~Clement, E.~L. Wong, L.~Alvisi, M.~Dahlin, and M.~Marchetti.
\newblock Making {Byzantine} fault tolerant systems tolerate {Byzantine}
  faults.
\newblock In {\em Proc.\ 6th Symp.\ Networked Systems Design and Implementation
  (NSDI)}, pages 153--168, 2009.

\bibitem{copzho14}
C.~Copeland and H.~Zhong.
\newblock Tangaroa: A {Byzantine} fault tolerant raft.
\newblock Class project in Distributed Systems, Stanford University,
  \url{http://www.scs.stanford.edu/14au-cs244b/labs/projects/copeland_zhong.pdf},
  Dec. 2014.

\bibitem{dmpz14}
S.~Duan, H.~Meling, S.~Peisert, and H.~Zhang.
\newblock {BChain}: {Byzantine} replication with high throughput and embedded
  reconfiguration.
\newblock In M.~K. Aguilera, L.~Querzoni, and M.~Shapiro, editors, {\em Proc.\
  18th Conference on Principles of Distributed Systems (OPODIS)}, volume 8878
  of {\em Lecture Notes in Computer Science}, pages 91--106. Springer, 2014.

\bibitem{dwlyst88}
C.~Dwork, N.~Lynch, and L.~Stockmeyer.
\newblock Consensus in the presence of partial synchrony.
\newblock {\em Journal of the ACM}, 35(2):288--323, 1988.

\bibitem{filypa85}
M.~J. Fischer, N.~A. Lynch, and M.~S. Paterson.
\newblock Impossibility of distributed consensus with one faulty process.
\newblock {\em Journal of the ACM}, 32(2):374--382, Apr. 1985.

\bibitem{gartne16}
D.~Furlonger and R.~Valdes.
\newblock Hype cycle for blockchain technologies and the programmable economy,
  2016.
\newblock
  \url{http://www.gartner.com/smarterwithgartner/3-trends-appear-in-the-gartner-hype-cycle-for-emerging-technologies-2016},
  July 2016.

\bibitem{gakile15}
J.~A. Garay, A.~Kiayias, and N.~Leonardos.
\newblock The bitcoin backbone protocol: Analysis and applications.
\newblock In {\em Advances in Cryptology: Eurocrypt 2015}, volume 9057 of {\em
  Lecture Notes in Computer Science}, pages 281--310. Springer, 2015.

\bibitem{greens16}
G.~Greenspan.
\newblock Multichain private blockchain~--- white paper.
\newblock \url{http://www.multichain.com/download/MultiChain-White-Paper.pdf},
  2016.

\bibitem{glpq10}
R.~Guerraoui, R.~R. Levy, B.~Pochon, and V.~Qu{\'{e}}ma.
\newblock Throughput optimal total order broadcast for cluster environments.
\newblock {\em ACM Transactions on Computer Systems}, 28(2):5:1--5:32, 2010.

\bibitem{hadtou93}
V.~Hadzilacos and S.~Toueg.
\newblock Fault-tolerant broadcasts and related problems.
\newblock In S.~J. Mullender, editor, {\em Distributed Systems (2nd Ed.)}. ACM
  Press \& Addison-Wesley, New York, 1993.
\newblock Expanded version appears as Technical Report TR94-1425, Department of
  Computer Science, Cornell University, Ithaca NY, 1994.

\bibitem{hearn16}
M.~Hearn.
\newblock Corda: A distributed ledger.
\newblock Available online,
  \url{https://docs.corda.net/_static/corda-technical-whitepaper.pdf}, 2016.

\bibitem{hkjr10}
P.~Hunt, M.~Konar, F.~P. Junqueira, and B.~Reed.
\newblock {ZooKeeper}: Wait-free coordination for internet-scale systems.
\newblock In {\em Proc.\ USENIX Annual Technical Conference}, 2010.

\bibitem{jurese11}
F.~Junqueira, B.~Reed, and M.~Serafini.
\newblock Zab: High-performance broadcast for primary-backup systems.
\newblock In {\em Proc.\ 41st International Conference on Dependable Systems
  and Networks}, 2011.

\bibitem{kbcdkm12}
R.~Kapitza, J.~Behl, C.~Cachin, T.~Distler, S.~Kuhnle, S.~V. Mohammadi,
  W.~Schr\"oder-Preikschat, and K.~Stengel.
\newblock {CheapBFT}: Resource-efficient {Byzantine} fault tolerance.
\newblock In {\em Proc.\ 7th European Conference on Computer Systems
  (EuroSys)}, pages 295--308, Apr. 2012.

\bibitem{Kleppmann17}
M.~Kleppmann.
\newblock {\em Designing Data-Intensive Applications}.
\newblock O'Reilly Media, 2017.

\bibitem{lampor98}
L.~Lamport.
\newblock The part-time parliament.
\newblock {\em ACM Transactions on Computer Systems}, 16(2):133--169, May 1998.

\bibitem{lampor01}
L.~Lamport.
\newblock Paxos made simple.
\newblock {\em SIGACT News}, 32(4):51--58, 2001.

\bibitem{lashpe82}
L.~Lamport, R.~Shostak, and M.~Pease.
\newblock The {Byzantine} generals problem.
\newblock {\em ACM Transactions on Programming Languages and Systems},
  4(3):382--401, July 1982.

\bibitem{lampso01}
B.~Lampson.
\newblock The {ABCD's} of {Paxos}.
\newblock In {\em Proc.\ 20th ACM Symposium on Principles of Distributed
  Computing (PODC)}, 2001.

\bibitem{lesozo15}
Y.~Lewenberg, Y.~Sompolinsky, and A.~Zohar.
\newblock Inclusive block chain protocols.
\newblock In R.~B{\"{o}}hme and T.~Okamoto, editors, {\em Proc.\ Financial
  Cryptography and Data Security (FC 2015)}, volume 8975 of {\em Lecture Notes
  in Computer Science}, pages 528--547. Springer, 2015.

\bibitem{liskov10}
B.~Liskov.
\newblock From viewstamped replication to {Byzantine} fault tolerance.
\newblock In B.~Charron-Bost, F.~Pedone, and A.~Schiper, editors, {\em
  Replication: Theory and Practice}, volume 5959 of {\em Lecture Notes in
  Computer Science}, pages 121--149. Springer, 2010.

\bibitem{liscow12}
B.~Liskov and J.~Cowling.
\newblock Viewstamped replication revisited.
\newblock MIT-CSAIL-TR-2012-021, July 2012.

\bibitem{malrei98a}
D.~Malkhi and M.~K. Reiter.
\newblock {Byzantine} quorum systems.
\newblock {\em Distributed Computing}, 11(4):203--213, 1998.

\bibitem{marewo00}
D.~Malkhi, M.~K. Reiter, and A.~Wool.
\newblock The load and availability of {Byzantine} quorum systems.
\newblock {\em SIAM Journal on Computing}, 29(6):1889--1906, 2000.

\bibitem{martin16}
W.~Martino.
\newblock Kadena~--- the first scalable, high performance private blockchain.
\newblock Whitepaper,
  \url{http://kadena.io/docs/Kadena-ConsensusWhitePaper-Aug2016.pdf}, 2016.

\bibitem{mazier16}
D.~Mazi{\`e}res.
\newblock The {Stellar} consensus protocol: A federated model for
  {Internet}-level consensus.
\newblock Stellar, available online,
  \url{https://www.stellar.org/papers/stellar-consensus-protocol.pdf}, 2016.

\bibitem{mxcss16}
A.~Miller, Y.~Xia, K.~Croman, E.~Shi, and D.~Song.
\newblock The honey badger of {BFT} protocols.
\newblock In {\em Proc.\ ACM Conference on Computer and Communications Security
  (CCS)}, 2016.

\bibitem{nakamo09}
S.~Nakamoto.
\newblock Bitcoin: A peer-to-peer electronic cash system.
\newblock Whitepaper, 2009.
\newblock \url{http://bitcoin.org/bitcoin.pdf}.

\bibitem{okilis88}
B.~M. Oki and B.~Liskov.
\newblock Viewstamped replication: A new primary copy method to support
  highly-available distributed systems.
\newblock In {\em Proc.\ 7th ACM Symposium on Principles of Distributed
  Computing (PODC)}, pages 8--17, 1988.

\bibitem{ongous14}
D.~Ongaro and J.~K. Ousterhout.
\newblock In search of an understandable consensus algorithm.
\newblock In {\em Proc.\ USENIX Annual Technical Conference}, pages 305--319,
  2014.

\bibitem{peshla80}
M.~Pease, R.~Shostak, and L.~Lamport.
\newblock Reaching agreement in the presence of faults.
\newblock {\em Journal of the ACM}, 27(2):228--234, Apr. 1980.

\bibitem{popov16}
S.~Popov.
\newblock The tangle.
\newblock White paper, available at \url{https://iota.org/IOTA_Whitepaper.pdf},
  2016.

\bibitem{Raynal10b}
M.~Raynal.
\newblock {\em Communication and Agreement Abstractions for Fault-Tolerant
  Asynchronous Distributed Systems}.
\newblock Synthesis Lectures on Distributed Computing Theory. Morgan \&
  Claypool, 2010.

\bibitem{ripple17}
Ripple.
\newblock Operating rippled servers.
\newblock Available online,
  \url{https://ripple.com/build/rippled-setup/#properties-of-a-good-validator},
  2017.

\bibitem{samman16}
G.~Samman.
\newblock Kadena: The first real private blockchain.
\newblock
  \url{http://sammantics.com/blog/2016/11/29/kadena-the-first-real-private-blockchain},
  Nov. 2016.

\bibitem{scbl09}
G.~{Santos~Veronese}, M.~Correia, A.~N. Bessani, and L.~C. Lung.
\newblock Spin one's wheels? {Byzantine} fault tolerance with a spinning
  primary.
\newblock In {\em Proc.\ 28th Symposium on Reliable Distributed Systems
  (SRDS)}, pages 135--144, 2009.

\bibitem{schnei90}
F.~B. Schneider.
\newblock Implementing fault-tolerant services using the state machine
  approach: A tutorial.
\newblock {\em ACM Computing Surveys}, 22(4):299--319, Dec. 1990.

\bibitem{schnei99}
B.~Schneier.
\newblock Snake oil.
\newblock \url{https://www.schneier.com/crypto-gram/archives/1999/0215.html},
  Feb. 1999.

\bibitem{scyobr14}
D.~Schwartz, N.~Youngs, and A.~Britto.
\newblock The {Ripple} protocol consensus algorithm.
\newblock Ripple Inc., available online,
  \url{https://ripple.com/files/ripple_consensus_whitepaper.pdf}, 2017.

\bibitem{swgmm17}
M.~Schwarz, S.~Weiser, D.~Gruss, C.~Maurice, and S.~Mangard.
\newblock Malware guard extension: Using {SGX} to conceal cache attacks.
\newblock e-print, arXiv:1702.08719 [cs.CR], 2017.

\bibitem{srmj12}
A.~Shraer, B.~Reed, D.~Malkhi, and F.~P. Junqueira.
\newblock Dynamic reconfiguration of primary/backup clusters.
\newblock In {\em Proc.\ USENIX Annual Technical Conference}, pages 425--437,
  2012.

\bibitem{soubes15}
J.~Sousa and A.~Bessani.
\newblock Separating the {WHEAT} from the chaff: An empirical design for
  geo-replicated state machines.
\newblock In {\em Proc.\ 34th Symposium on Reliable Distributed Systems
  (SRDS)}, pages 146--155, 2015.

\bibitem{sritou87}
T.~K. Srikanth and S.~Toueg.
\newblock Simulating authenticated broadcasts to derive simple fault-tolerant
  algorithms.
\newblock {\em Distributed Computing}, 2:80--94, 1987.

\bibitem{swanso15}
T.~Swanson.
\newblock Consensus-as-a-service: A brief report on the emergence of
  permissioned, distributed ledger systems.
\newblock Report, available online, Apr. 2015.
\newblock URL:
  \url{http://www.ofnumbers.com/wp-content/uploads/2015/04/Permissioned-distributed-ledgers.pdf}.

\bibitem{rescsc15}
R.~van Renesse, N.~Schiper, and F.~B. Schneider.
\newblock Vive la diff{\'{e}}rence: Paxos vs. viewstamped replication vs. zab.
\newblock {\em IEEE Transactions on Dependable and Secure Computing},
  12(4):472--484, 2015.

\bibitem{rensch04}
R.~van Renesse and F.~B. Schneider.
\newblock Chain replication for supporting high throughput and availability.
\newblock In {\em Proc.\ 6th Symp.\ Operating Systems Design and Implementation
  (OSDI)}, 2004.

\bibitem{Vukolic12}
M.~Vukoli\'{c}.
\newblock {\em Quorum Systems: With Applications to Storage and Consensus}.
\newblock Synthesis Lectures on Distributed Computing Theory. Morgan \&
  Claypool, 2012.

\bibitem{vukoli17}
M.~Vukoli{\'c}.
\newblock Rethinking permissioned blockchains.
\newblock In {\em Proc.\ ACM Workshop on Blockchain, Cryptocurrencies and
  Contracts (BCC '17)}, 2017.

\end{thebibliography}
\end{document}